\documentstyle[epsfig,longtable]{aipproc}

\begin{document}

\title{Two-dimensional arrays of low capacitance tunnel junctions:\\
general properties, phase transitions and Hall effect}

\author{P. Delsing$^*$, C.D. Chen$^{\dagger *}$, D.B Haviland$^*$, T. Bergsten$^*$\\
and T. Claeson$^*$}
\address{$^*$Department of Microelectronics and Nanoscience,\\
Chalmers University of Technology and G\"{o}teborg University, S-412 96, G\"oteborg Sweden\\
$^{\dagger}$Institute of Physics, Academia Sinica, Nankang, Taipei 11529, Taiwan }

\maketitle


\begin{abstract}
We describe transport properties of two-dimensional 
arrays of low 
capacitance tunnel junctions, such as the current voltage 
characteristic and its dependence
on external magnetic field and temperature. We discuss
several experiments in which the small capacitance of the junctions plays
an important role, and we also describe the methods for
fabrication and measurements. 

In arrays where the junctions have a relatively large charging 
energy, ($i.e.$ when they have a low capacitance) and 
a high normal state resistance, the low bias resistance increases with
decreasing temperature and eventually at very low temperature the whole array may 
become insulating even though the electrodes in the array are 
superconducting. This transition to the insulating state can be described 
by thermal activation, characterized by an activation energy. We find 
that for certain junction parameters the activation energy
oscillates with magnetic field with 
a period corresponding to one flux quantum per unit cell.

In an intermediate region where the junction resistance is of the order
of the quantum resistance and the charging energy is of the order of
the Josephson coupling energy, the arrays can be tuned between a 
superconducting and an insulating state with a
magnetic field. We describe measurements of this magnetic-field-tuned
superconductor insulator
transition, and we show that the resistance data can be scaled over 
several orders of magnitude. Four arrays follow the same universal 
functin provided we use a modified scaling parameter. We find a critical 
exponent close to unity, in good agreement with the theory.

At the transition the transverse (Hall) resistance is found to
be very small in comparison with the longitudinal resistance.
However, for magnetic field values larger than the critical value.
we observe a substantial Hall resistance.
The Hall resistance of these arrays oscillates with the applied 
magnetic field. Features in the magnetic field dependence of the 
Hall resistance can qualitatively be correlated to features in the 
derivative of the longitudinal resistance, similar to what is found in 
the quantum Hall effect.
\end{abstract}


\section{Introduction}
\label{Introduction}
The electrical transport in two-dimensional (2D) arrays\index{2D-arrays}
of small Josephson 
tunnel junctions\index{Josephson junction} can be described either in terms of charges or in 
terms of vortices \cite{Mooij/Schon-LH,Fazio/Schon}.
Transport of charge generates a current which acts as a driving force for the 
vortices. The charge transport may be obstructed by a large charging 
energy, $E_C\equiv e^2/2C$, $C$ being the capacitance of the individual junction. 
On the other hand, transport of vortices generates a voltage which acts as a driving 
force for the charges. The vortex transport may be hindered if the Josephson 
coupling  energy\index{Josephson coupling  energy}
is large. At low temperature the Josephson 
coupling  energy is given by $E_{J}\equiv (R_{Q}/R_{N})/(\Delta /2)$,
where $R_{N}$ is the normal state resistance of the individual 
junctions, $R_{Q}\equiv h/4e^2\approx 6.45\,$k$\Omega$ is the quantum resistance
\index{quantum resistance},
and $\Delta$ is the superconducting energy gap of the electrodes.

\begin{figure}[b!]
\centerline{\epsfig{file=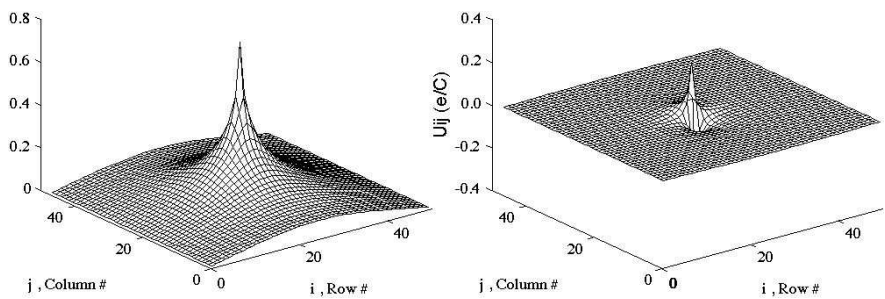,width=12cm}}
\vspace{10pt}
\caption{\label{soliton}The exact solution for the potential distribution $U_{ij}$ in a $49\times 47$ 
junction array, were the ratio between the junction capacitance and the self 
capacitance of each electrode is 400. The current leads are connected at 
$i=1$ and $i=49$  a) A single electron in the center of the array giving rise to a 
single electron soliton. b) The fundamental excitation, a soliton/anti-soliton 
pair at neighboring sites.}
\end{figure}

Adding a single charge to an electrode in the array gives rise 
to an electrostatic potential distribution which is sometimes referred 
to as a charge soliton\cite{Bakhvalov-2D}\index{charge soliton}. A missing charge gives
rise to the counterpart, an anti-soliton. The 
charge solitons can move in the array by single-charge tunnel events.
When the electrodes are superconducting, both Single Electron Solitons (SES) and 
Cooper Pair Solitons (CPS) can exist. An exact calculation of the potential 
distribution in a $49\times 47$ junction array assuming only nearest
neighbor coupling, is shown in 
Fig.\,\ref{soliton}, for two different 
cases: A single electron in the center of the array, giving rise to a SES, and the 
fundamental excitation, a soliton/anti-soliton pair at adjacent electrodes.
The size of the soliton is given by $\sqrt{C/C_{0}}$ as long as 
$C>C_{0}$. $C_{0}$ is the capacitance between an electrode in the 
array and infinity. However if $C$ is of the order of $C_{0}$ the nearest
neighbor approach is not a good approximation \cite{Orlando-APL96}.

On the other hand, due to the existence of a macroscopic phase of the
superconductor, a vortex\index{vortex}, {\it i.e.} a phase winding of 2$\pi$, with an 
associated magnetic flux quantum $\Phi_0\equiv h/2e$, can exist in a loop
of the array.
In a two-dimensional (2D) array of small Josephson tunnel junctions 
the duality between charges and vortices is especially pronounced. 
Therefore it is a very suitable object to study the dynamics of both
charges and vortices. 

The arrays which have high junction resistance, $R_N\gg R_Q$
and large charging energy, $E_C\gg E_J$, become insulating at low temperature 
due to the Coulomb 
blockade\cite{Averin/Likharev,LesHouches}\index{Coulomb blockade}, 
regardless of whether the electrodes are superconducting or normal.
 
In the opposite limit, $R_N\ll R_Q$ and $E_C\ll E_J$, superconductivity prevails 
and the resistance goes to zero at low temperature. In this limit where the Josephson
energy dominates, 
vortices act as classical or quasi-classical particles and there is a large number of 
papers describing such systems theoretically 
\cite{Mooij-NATO86,Mooij/Schon,Frascati-95,Minnhagen-RMP}
as well as experimentally
\cite{Voss/Webb,LAT,Wees-PRB87,Tighe-PRB91,vdZant-PRB93,Rzchowski-PRB94,CCD-PRB96}
 With normal electrodes and 
$R_N\gg R_Q$ the array goes insulating at low temperature while in the opposite 
limit $R_N\ll R_Q$, the array stays resistive even down to very low temperature.

In this paper we will concentrate on the arrays where the charging 
energy\index{charging energy} is comparable to or larger than the Josephson coupling
energy, and where
the junction resistance is of the order of or larger than $R_{Q}$.
Throughout this paper, we will refer to an array with normal electrodes as 
being in the N-state, and to an array with superconducting electrodes as being 
in the S-state.

If the interaction between the charges (or vortices) is logarithmic, the 
transition to the insulating (superconducting) state would be of the 
Kosterlitz-Thouless-Berezinskii (KTB) type\cite{KT(B),(KT)B}\index{KTB 
transition}. 
As the temperature is increased, an 
insulating 2D array can undergo a charge unbinding 
transition\cite{Mooij-PRL90} at some 
transition temperature, which results in a conductive state. Likewise, a 
superconducting array can undergo a vortex unbinding transition to a resistive 
state. It has been shown that the superconducting transition 
can be described as a vortex unbinding, KTB transition\cite{Voss/Webb}. 
In more recent years there has also been a lot of interest in arrays
where the dynamics is best described in terms of charges
\cite{Mooij-PRL90,GeerligsPhysicaB,Geerligs-PRL89,CCD-PhysicaScripta,Delsing-SQUID91,Tighe-PRB93,Delsing-PRB94,Kanda/Kobayashi}. 
In several papers there has been a discussion whether the observed 
transition to the insulating state can be described as a KTB transition as 
well\cite{Mooij-PRL90,Delsing-SQUID91,Tighe-PRB93}. 
In a paper by Tighe et al.\cite{Tighe-PRB93} it was pointed out that the transition 
could be well described by thermal activation\index{thermal activation} of charge solitons.
Comparing 
results of three different 
groups\cite{Mooij-PRL90,Delsing-SQUID91,Tighe-PRB93}, 
they found an activation energy 
$E_a\approx \frac{1}{4}E_C$ in the N-state, a value which can be theoretically justified. 
They also suggested that in the S-state at $B=0$, $E_a$ should be 
$\frac{1}{4}E_C+\Delta _0$, where $2\Delta _0$ is the 
superconducting energy gap at $B=0$ and $T=0$. An interesting new 
result is that of Kanda and Kobayashi \cite{Kanda/Kobayashi} where they find a thermal 
activation behavior at higher temperatures but a stronger dependence 
at lower temperature in the N-state. An interesting theoretical development in 
this field is a recent paper by Feigelman et al.\cite{Feigelman}, where they treat 
the posibility of parity effects 2D arrays. In section
\ref{Insulating} we will describe measurements where 
we have investigated the transition to the insulating state extensively.

For arrays in the intermediate regime ($R_N\approx R_Q$ and 
$E_C\approx E_J$) which just barely go superconducting,
a small magnetic field can drastically change the low bias resistance 
and in fact drive the array into the insulating state \cite{vdZant-PRL92}. In 
a theoretical description of this effect \cite{Fischer-PRL90}, the field induced 
excess vortices drive the system from a vortex glass superconducting 
phase into a Bose-condensed insulating phase. The zero-magnetic-field 
KTB vortex-unbinding transition is replaced 
by a field-tuned vortex-delocalization transition.
A superconductor-insulator(SI) transition\index{superconductor insultor transition}
can also be driven by other 
external variables such as electric field \cite{Bruder,Lafarge-95} or 
dissipation \cite{Rimberg-PRL97}. In Section \ref{SIT}
we show experiments on the magnetic-field-tuned superconductor-insulator
transition. The zero bias resistance, 
$R_{o}$ was measured as a function of temperature and frustration.
The frustration, $f$ is defined as the magnetic
field normalized to $B_{o}$, 
where $B_{o}$ is the field corresponding to one flux quantum 
per unit cell in the array. The scaling curves demonstrate how $R_{o}$, plotted 
as a particular function of $T$ and $f$, display a transition
from insulator to superconductor.
According to theory the resistance at the critical frustration $f_{c}$
should be universal and equal to $R_{Q}$. 
From the data of four different arrays we find a value which is of the 
order of $R_{Q}$ but sample dependent. We can also deduce a dynamic exponent close 
to unity, which is in agreement with the theory \cite{Fischer-PRL90}.

Right at the critical frustration, the Hall resistance\index{Hall resistance}
is very small compared with the longitudinal resistance, indicating 
a small Hall effect at the SI transition. However for frustration 
values larger than the critical frustration the Hall resistance can 
be relatively large. Hall measurements in both conventional 
superconductors \cite{Graybeal-PRB94,Smith-PRB94} and 
high-$T_c$ superconductors \cite{Samoilov-PRB94,Harris-PRB94}
have shown a sign reversal of the 
Hall resistivity in the vicinity of the superconducting transition 
temperature $T_c$, where the samples are in a mixed state. Hall 
measurements have also been performed on disordered superconductors 
near the superconductor-to-insulator transition \cite{Paalanen-PRL92}. 
In Section \ref{Hall} we present the frustration dependence
of the longitudinal resistance $R_{0xx}$ and the Hall resistance $R_{0xy}$.
Both the longitudinal and Hall resistance are periodic functions of the
magnetic filed, and the Hall resistance changes sign at several magnetic 
fields within one period.
We also describe the dc measurements of longitudinal voltage 
$V_x$ and the Hall voltage $V_y$ as a function of bias current $I_x$.


\section{Sample Fabrication and Measurements}
\label{Fab}
The arrays are fabricated\index{fabrication} on unoxidized silicon substrates, using a 
combination of photo- and
electron-beam-lithography and an angle evaporation technique. 
Aluminum is used for both top and bottom electrodes. 
The number of junctions in each row, 
N, is the same as the number of junctions in each column, with N ranging 
between 10 and 168. Therefore, the array resistance equals the individual 
junction resistance, assuming a homogeneous array.

The samples are made in two steps. First a gold contact pattern is 
made with conventional photo lithography, and then the actual array is
made with electron-beam lithography 
The contact pattern contains a large number of 7x7\,mm$^2$ chips 
distributed over a 2 inch wafer area, each with 16 
contact pads leading to a central area of 160x160\,$\mu$m$^2$.
A double metal layer of 20\,nm chromium-nickel and 80\,nm gold is evaporated
and the redundant metal is lifted off in acetone. 
The chromium-nickel film makes the gold stick 
better to the surface.

A double layer e-beam resist consisting of a $\sim$210\,nm thick 
bottom layer of P(MMA/MAA) copolymer is spun onto the wafer and a
$\sim$60\,nm thick top layer of 
PMMA(950k) is used. The chip is mounted in an e-beam 
lithography instrument and the central area of each chip is 
exposed using the array pattern. A current of 20-30\,pA 
corresponding to a beam size of about 10\,nm is used. The beam voltage 
is 50\,kV and the area dose is 160-200\,$\mu$C/cm$^2$.

Each chip is then developed 
in two different developers: first for $\sim$10-20\,s in the PMMA 
developer which consists of a 1:3 mixture of toluene and isopropanol, then  
for 20-40\,s in the copolymer developer which consists of a 1:5 mixture of 
ethyl-cellosolve-acetate (ECA) and ethanol. As an alternative the nontoxic
mixture of 5-10 \% water in isopropanol can be used to develop both 
layers in one step, with a development time of 1-2 minutes.

After development, the resist mask contains an undercut pattern with suspended 
bridges\cite{Niemeyer,Dolan} which will be used to form the junctions.
By depositing bottom 
and top electrodes from different angles the overlap can be controlled. 
The base and top electrodes are evaporated from tungsten boats, while 
the substrate holder is tilted at two different angles
($\sim \pm 15\,^{o}C$) to give the 
desired overlap. Before the top electrode is deposited, a tunnel 
barrier is formed by introducing 0.01-0.1\,mbar of oxygen to the chamber for 
3-10 minutes, by adjusting the oxidation parameters we get
the desired junction resistance. 

\begin{figure}[b!]
\centerline{\epsfig{file=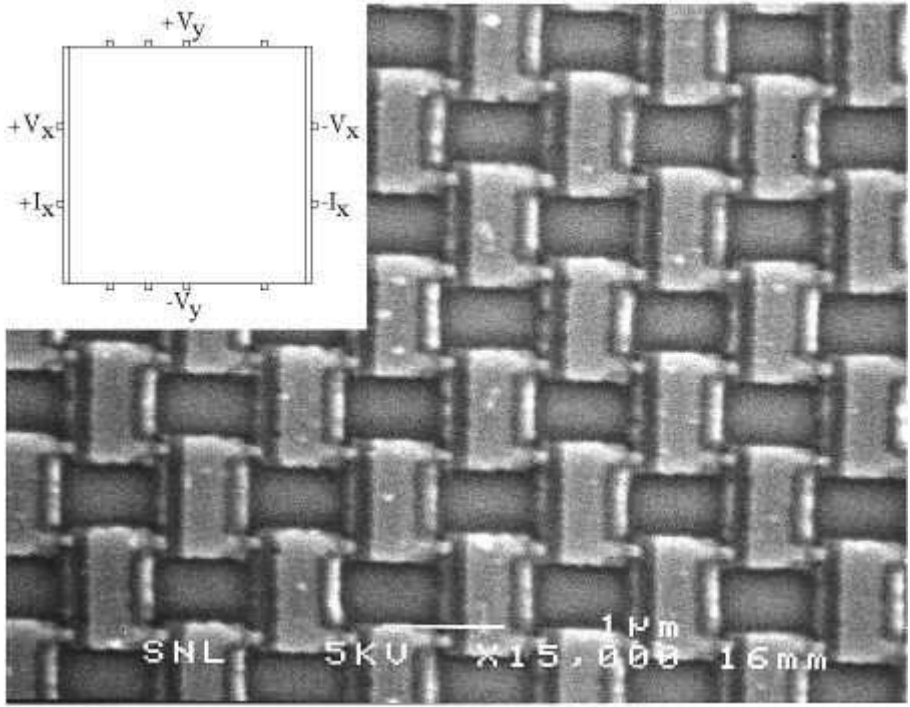,width=12cm}}
\vspace{10pt}
\caption{\label{Layout}A SEM picture of a part of an array.  Each island measures 
0.5 by 1 $\mu$m, the overlap between neighboring islands defines the 
tunnel junction. The insert shows the probe layout.  Each Hall probe 
consists of $3\times 3$ junctions, the junctions are identical to the 
junctions in the array.}
\end{figure}

A drawback with the angle evaporation technique is that a relatively 
large junction is formed in series with the small tunnel junction, see 
Fig.\,\ref{Layout}. 
The effect of this larger junction can in most cases be neglected, if its area is 
much larger than the area of the smaller junction. This is easy to 
make, but this requirement limits the minimum unit cell size of the 
array, which in turn decreases the soliton size.

The fabrication procedure results in junctions with normal state 
resistances in the range 4 to 150\,k$\Omega$, capacitances
of the order of 1\,fF, and $\Delta_0\approx 200$\,$\mu$V per 
junction. The typical unit cell size is of the order of 
$A_{cell}\approx 1\,\mu$m$^2$
These values are deduced from the $IV$-characteristics, assuming 
an offset voltage of $V_{off}=Ne/2C$
(for a discussion of the offset value see Ref.\cite{Tighe-PRB93}). 
The Josephson coupling energies of the individual junctions are 
determined\cite{Ambegaokar/Baratoff} from $R_N$ and $\Delta _0$. The 
superconducting transition temperature $T_{c}$ for the aluminum is in
the range of 1.35 to 1.60\,K.

The self capacitance of each electrode $C_0$ depends 
on the size of the electrode and the size of the array, and can be estimated to be of the 
order 10-20\,aF for the smaller arrays (\#2 and \#3) and about 2\,aF 
for the other arrays. This results in a soliton size in the range of 11 to 35, 
measured in units of the lattice spacing. 

The most important parameters of the 2D arrays described in this paper are
listed in Table \ref{Table1}. The arrays are divided into 3 groups.
Arrays \#1-3 show a 
decreasing resistance for decreasing temperature, and are referred to as the 
"superconducting" arrays. They have a low resistance, $R_N<R_Q$ and a 
relatively large Josephson coupling energy, $E_J/E_C>1$. 

Arrays \#4-8 show also go superconducting at low temperature, but they
display the magnetic field tuned SI transition, and are referred to as the 
"intermediate" arrays. They have $R_N\approx R_Q$ and $E_J/E_C\approx 1$.

Arrays \#9-15 
show an increasing resistance for decreasing temperature, and are referred to as the 
"insulating" arrays. They have a high resistance, $R_N>15$\,k$\Omega$ and a relatively 
small Josephson coupling energy, $E_J/E_C<0.5$.

\begin{table}[b!]
\caption{\label{Table1}Parameters for the 15 arrays. The resistance $R_N$, the capacitance $C$, 
and the superconducting energy gap 2$\Delta_0$, were deduced from the $IV$-curves.  
The charging energy $E_C$, and the Josephson coupling energy $E_J$, were calculated 
from these values. $B_0$ is the magnetic field corresponding to one flux quantum per 
unit cell. $\Lambda$ is the soliton size measured in units of the lattice spacing. 
$E_{aN}$ is the activation energy for normal electrodes.}
\begin{center}
\begin{tabular}{|r|r|d|d|d|d|d|d|d|d|}
\hline
\# & $N$ & $R_N$ & $B_0$ & $\Lambda$ & $E_J/k_B$
& $E_C/k_B$ & $\Delta_0/E_C$ & $E_{aN}/E_C$ & $E_{J}/E_C$ \\
~ & ~ & (k$\Omega$) & (G) & ~ & (K)
& (K) & ~ & ~ & ~ \\
\hline
1 & 112 & 3.98 & 10.4 & 35 & 1.82 & 0.37 & 6.08 & 0.56 & 4.91 \\
2 & 20 & 4.08 & 10.4 & 16 & 1.78 & 0.36 & 6.12 & 0.66 & 4.93 \\
3 & 10 & 4.49 & 10.4 & 11 & 1.62 & 0.38 & 5.98 & 0.70 & 4.25 \\
~ & ~ & ~ & ~ & ~ & ~ & ~ & ~ & ~ & ~ \\
4 & 146 & 7.54 & 16.3 & 19 & 0.95 & 0.60 & 3.64 & - & 1.57 \\
5 & 168 & 10.7 & 16.3 & 23 & 0.68 & 0.55 & 3.97 & - & 1.24 \\
6 & 146 & 12.5 & 16.3 & 27 & 0.57 & 0.72 & 3.04 & - & 0.80 \\
7 & 168 & 13.5 & 16.3 & 27 & 0.53 & 0.59 & 3.71 & - & 0.90 \\
8 & 146 & 13.5 & 16.3 & 29 & 0.56 & 0.73 & 3.25 & 0.24 & 0.77 \\
~ & ~ & ~ & ~ & ~ & ~ & ~ & ~ & ~ & ~ \\
9 & 146 & 24.4 & 16.3 & 22 & 0.33 & 1.22 & 2.05 & 0.24 & 0.27 \\
10 & 80 & 35.4 & 27.6 & 23 & 0.21 & 0.88 & 2.62 & 0.25 & 0.24 \\
11 & 80 & 38.0 & 27.6 & 22 & 0.194 & 0.92 & 2.49 & 0.24 & 0.21 \\
12 & 100 & 49.3 & 43.1 & 24 & 0.159 & 1.27 & 1.91 & 0.26 & 0.12 \\
13 & 100 & 59.7 & 43.1 & 23 & 0.132 & 1.35 & 1.82 & 0.25 & 0.10 \\
14 & 80 & 88.4 & 27.6 & 21 & 0.090 & 1.09 & 2.27 & 0.28 & 0.08 \\
15 & 100 & 151. & 43.1 & 20 & 0.057 & 1.75 & 1.52 & 0.31 & 0.03 \\
\hline
\end{tabular}
\end{center}
\end{table}

\begin{figure}[t!]
\centerline{\epsfig{file=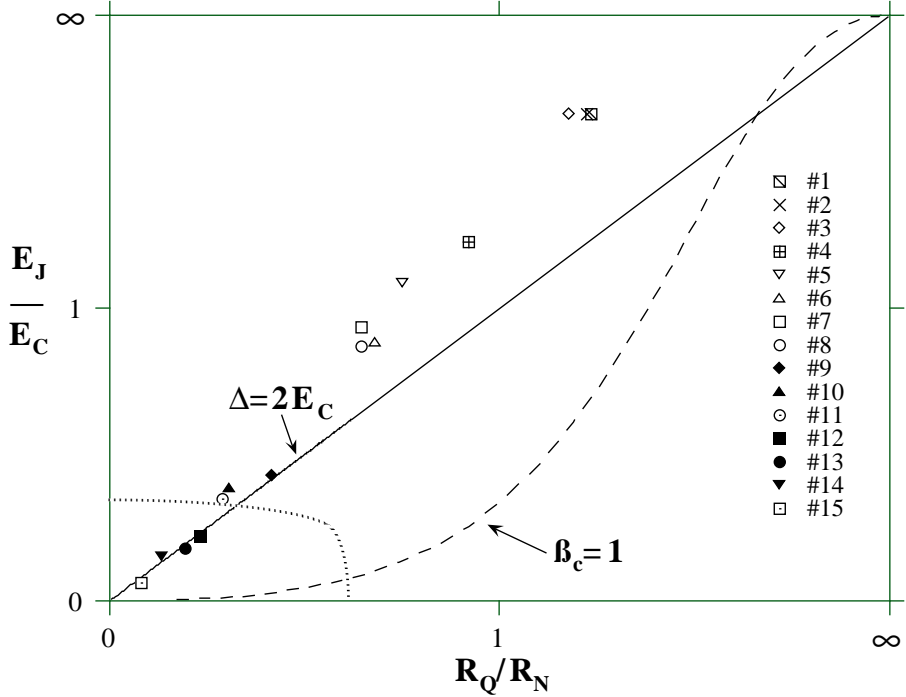,width=12cm}}
\vspace{10pt}
\caption{\label{Schmid}A $T=0$ phase diagram showing the $E_J/E_C$ and the $R_Q/R_N$ 
parameters for the measured arrays. The diagonal line represents 
$\Delta _0=2E_C$. The dotted line represents the border to the insulating region 
predicted by Fazio and Sch\"on. The dashed line 
corresponds to a Stewart MacCumber parameter of $\beta_{c}=1$.}
\end{figure}

The arrays are mapped onto a so called
quasi-Schmid diagram\cite{Schmid,quasiSchmid}\index{quasi-Schmid diagram}
in Fig.\,\ref{Schmid}, showing the $E_J/E_C$ and the $R_Q/R_N$ parameters of each array. 
The $E_J/E_C$ and $R_Q/R_N$ values in the range 0 to $\infty$ are scaled onto 
the horizontal and vertical axes in the range 0 to 2, using the function f(z)=2z/(1+z).
The diagonal line represents $\Delta_0=2E_C$, {\it i.e.} where the 
charging energy for a Cooper-pair is equal to 2$\Delta_{0}$.
The dotted line represents the border to the insulating region 
predicted by Fazio and Sch\"on \cite{Fazio/Schon}. The dashed line 
corresponds to a Stewart MacCumber parameter\index{Stewart MacCumber Parameter}
of $\beta_{c}\equiv (\pi^2/2)(E_J/E_C)(R_N/R_Q)^2=1$.
Above this line a classical 
Josephson junction (in the upper right part of the diagram) shows 
hysteresis \cite{Stewart,MacCumber}.

Our classification of the ``insulating'' arrays roughly 
agrees with the classification suggested by Fazio and Sch\"on\cite{Fazio/Schon} 
shown as a dashed line in Fig.\,\ref{Schmid}. However, a few of the arrays (\#9-11) 
which show "insulating" behavior fall slightly outside the insulating area predicted 
by Fazio and Sch\"on. 

In Section \ref{Insulating}, we will concentrate on the "insulating" arrays, 
but we will also present the N-state properties of the "superconducting" arrays.
The N-state may be thought of as lying on the x-axis of the diagram in 
Fig.\,\ref{Schmid}.
Sections \ref{SIT} and \ref{Hall} will discuss the ``intermediate'' arrays.

Fig.\,\ref{Layout} shows a typical sample, and the inset shows
the probe layout for the Hall measurements. There are four Hall probe pairs
situated at 1/6, 1/3, 1/2 and 
5/6 the distance between the ends of the array.
Each Hall probe is actually an array of 
$3\times 3$ junctions, with the outer side shorted by a strip from 
which the Hall voltage is taken.  In this way we can reduce the 
influence on the sample behavior arising from the presence of the 
voltage probes, and we are less sensitive to local defects possibly 
occurring in the probe. The resistance of the probe arrays is 
presumably similar to that of the array itself.  
This resistance is much smaller than the input impedance 
of our voltage amplifiers. 

The measurements are performed in a dilution refrigerator 
which is situated in an electrically shielded room.
A magnetic field up to about 1400\,G is applied perpendicular to the 
substrate. The magnetic field needed to produce one magnetic flux quantum per 
unit cell, $B_0$, varied between 10.4\,G and 43.1\,G for the different 
arrays. This is much less than the critical magnetic field, $B_c$ ($\sim 800$\,G), 
which was needed to bring the aluminum electrodes into the N-state.

All measurements are made with the biasing and the measurement 
circuitry symmetric with respect to ground. The current-voltage($IV$)-characteristics of the 
arrays are recorded at temperatures down to about 25\,mK.
The threshold voltages are deduced from the IVC measured
as the voltage at which the current has dropped to 
two times the noise level (going from large bias). The rms current-noise 
determined at low voltage is typically of the order of 0.1\,pA or less.

For the insulating arrays we deduce the zero bias resistance $R_0$
from the $IV$-characteristics.
For the ``superconducting'' and ``intermediate arrays'', $R_0$ can also be 
measured with an ac-technique using lock-in amplifiers, 
with excitation currents in the range 0.01-3 nA, and frequencies of 
the order of 10\,Hz.

The longitudinal voltage $V_x$ is measured at the 
superconducting strips at the ends of the array, and the Hall voltage 
$V_y$ is measured on the probes located on opposite sites of the 
array.  The Hall data presented here were 
measured from the probes located in the center of the array, see the 
inset of Fig.\,\ref{Layout}


\section{Current voltage characteristics}
\label{IVC}
The large scale $IV$-characteristics\index{current voltage characteristics}
of the arrays are very similar for the 
11 arrays and they resemble the $IV$-characteristics of a single high resistance 
Josephson junction. In the S-state there is a sharp rise in the conductance at a 
voltage $\sim N\cdot 2\Delta_0/e$. At high voltages the $IV$-characteristics are linear 
and there is the usual offset voltage which is due to the Coulomb 
blockade \cite{Averin/Likharev,LesHouches}. In the N-state the gap feature disappears 
but the offset voltage remains.

For low bias in the S-state, the $IV$-characteristics differ drastically for the different 
arrays. Qualitative similarities can be found between these $IV$-curves and those of 
single junctions biased through high impedance resistors \cite{Haviland-ZPB}. 
Arrays \#1-8 show a supercurrent-like feature at low bias, and $R_0$ decreases for 
decreasing temperature.
Arrays \#9-15 show a Coulomb blockade feature at low bias, and $R_0$ increases for 
decreasing temperature.

\subsection{The threshold voltage}
\label{Threshold}

The threshold voltage $V_t$\index{threshold voltage}
is the voltage at which solitons can be injected into the array.
According to theory \cite{Bakhvalov-2D}, the threshold voltage for injection of 
SESs in the N-state should be

\begin{equation}
\label{Vteq}
V_{tN}=2\left(1-\frac {2}{\pi}\right)\frac{E_C}{e}\Lambda
\end{equation}

for a symmetrically biased array with $C\gg C_0$. It is important to make the 
distinction between  symmetric and asymmetric (one side grounded) bias of the 
array, since the latter gives a factor of two lower threshold voltage.
However, it has been show by Middelton and Wingreen
\cite{Middelton/Wingreen} that the background charges modifies the 
picture and that the threshold voltage actually scales with the length 
of the array if the effect of random background charges is taken into 
account. 

\begin{figure}[t!]
\centerline{\epsfig{file=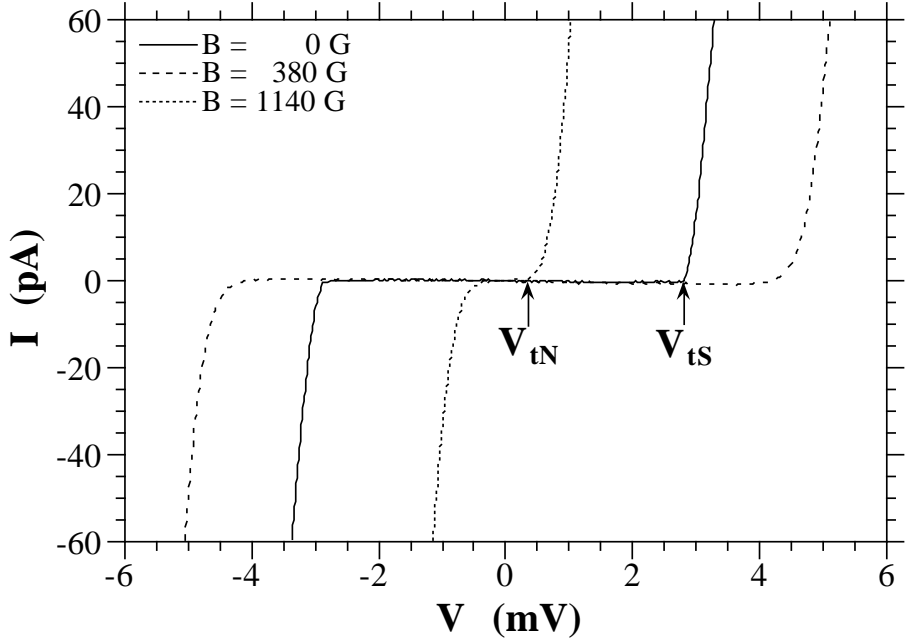,width=12cm}}
\vspace{10pt}
\caption{\label{Vt}The $IV$-characteristics of sample \#15 at different values of magnetic 
field. In the S-state (B=0\,G), the threshold is $V_{tS}=2.75$\,mV, while in the 
N-state ($B=1140$\,G), the threshold is $V_{tN}=0.25$\,mV. Note that the onset of 
current occurs at an even larger voltage for an intermediate value of $B$.}
\end{figure}

In the N-state, all arrays show Coulomb blockade feature at low temperature and low bias, 
but the threshold is smeared. However there is no sharp 
threshold voltage in the N-state for most of the arrays. 
Only for array \#15 , which has the largest $R_N$, we can deduce a 
threshold voltage of about 0.25\,mV(see Fig.\,\ref{Vt}). 
Note that the onset of current occurs at a 
substantially higher voltage for intermediate magnetic fields. For those fields 
however the onset was more gradual. A similar behavior was observed in all 
of the "insulating" arrays.

All of our arrays were symmetrically biased and for array \#15 we get a 
theoretical value of $V_{tN}=2.3$\,mV, according to Eq.\,\ref{Vteq}.
The fact that the measured value is 
substantially lower than the theoretically predicted one, is consistent with the 
picture that quantum fluctuations effectively lower the energy barrier for 
injection of charge \cite{Delsing-PRB94}.

\begin{figure}[t!]
\centerline{\epsfig{file=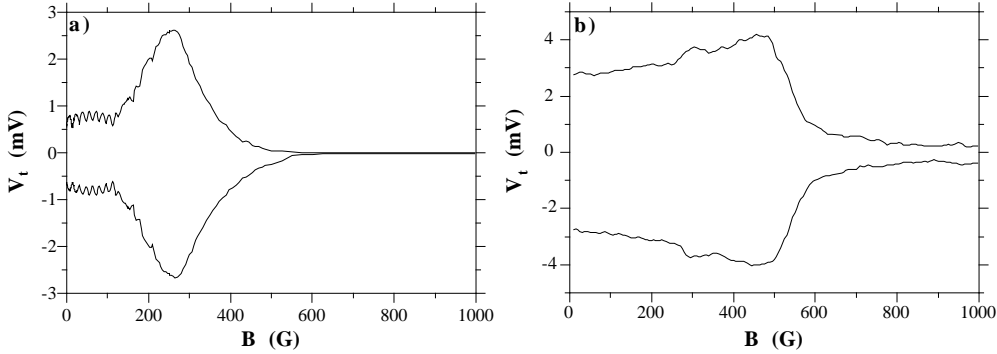,width=14cm}}
\vspace{10pt}
\caption{\label{Vt(B)}The positive and negative threshold voltages as a function of $B$ for 
two of the arrays at $T\approx 25$\,mK,  a) \#9  At low 
magnetic field $V_t$ oscillates with one flux quantum per unit cell, showing 
that Cooper pair solitons are injected. b) \#15.}
\end{figure}

In the S-state there was a sharp threshold for all the 
"insulating" arrays, except for arrays \#10 and \#11.
$V_t$ is shown as a function of $B$ for four of the arrays in Fig.\,\ref{Vt(B)}.
For several of the arrays $V_t$ oscillates with $B$, demonstrating that 
Cooper pair solitons are injected at low magnetic fields. The period of 
oscillation corresponds to one flux quantum per unit cell and agrees 
well with the $B_0$ values of the different arrays. The oscillations in 
$V_t$ correspond to the oscillations in activation energy, described in Section 
\ref{Insulating}, and the 
oscillation peaks in $V_t$ and $E_a$ occur at the same $B$ values 
(compare Figs.\,\ref{Ea(B)} and \ref{Vt(B)}).
For increasing magnetic field the threshold voltage increases and peaks at a 
field in the range of 250 to 450\,G, which is well below the critical field $B_c$ for 
the electrodes. $V_t$ then decreases rapidly at larger $B$.

The threshold voltage in the S-state has, to our knowledge, not been 
described theoretically in the literature. 
However, following the arguments in Ref.\cite{Bakhvalov-2D}, 
and neglecting the Josephson 
coupling energy, we can estimate $V_t$ for two different situations. For direct 
injection of Cooper pair solitons, we get a threshold voltage which is two times 
higher than the value for single electron solitons, $V_{tS2e}=2V_{tN}$. The other 
possibility is that a Cooper-pair is broken up and then a single electron is 
injected. Then the threshold would be
$V_{tSe}=V_{tN}+\Lambda 4\Delta_{0} /e$ 
for the case of symmetric bias. 
For all our arrays $V_{tS2e}<V_{tSe}$ at $B=0$, and therefore we 
would expect injection of Cooper pairs to be responsible for the threshold.

This agrees well with our observation of the oscillating $V_t$ at low magnetic 
field. The increase of $V_t$ with increasing magnetic field can be understood in 
the following way. Since $\Delta $, and thereby $E_J$, decreases with increasing magnetic field it 
becomes harder to inject Cooper-pairs, and therefore $V_t$ increases. 
At some magnetic field the threshold 
for single electrons will become equal to that for 
Cooper-pairs ($V_{tS2e}=V_{tSe}$) 
and we would expect a crossover from Cooper-pair  injection to single 
electron injection. This is observed in all the samples as a peak in $V_t$ at 
magnetic fields in the range 250 to 450\,G. 
Beyond this crossover $V_{tSe}$ decreases 
as a function of increasing $B$, due to the decreasing $\Delta $.

To get a more quantitative description, other effects depending on the array size and the 
background charge, as well as co-tunneling and the Josephson coupling, 
would have to be included. The fact that the observed values at $B=0$ are 
generally lower than the "theoretical" value $V_{tS2e}$, can possibly be explained 
if the Josephson coupling energy is taken into account. A step in this 
direction is a recent paper on 1D-arrays where the threshold 
dependence on $E_{J}$ is discussed \cite{Haviland/Delsing}. The 
current above the threshold has been analyzed using scaling theory by 
Rimberg {et al.} \cite{Rimberg-PRL95}.

\begin{figure}[b!]
\centerline{\epsfig{file=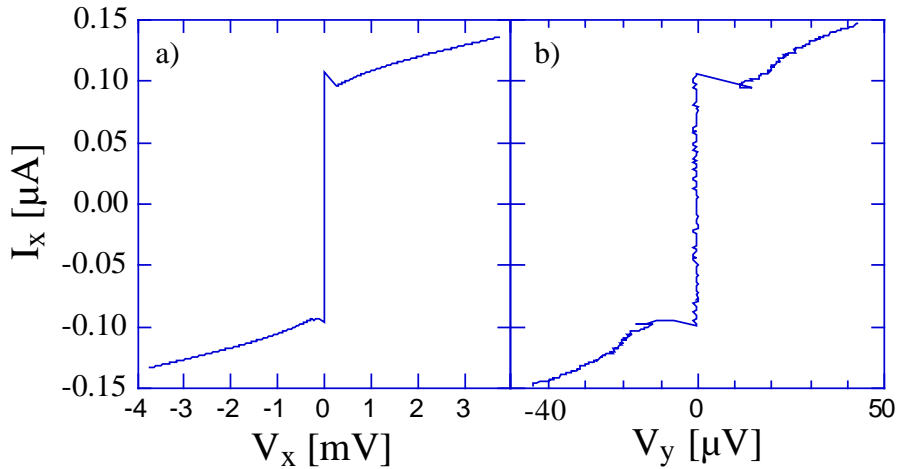,width=12cm}}
\vspace{10pt}
\caption{\label{IV}The longitudinal voltage $V_x$ (a), and the Hall 
voltage $V_y$ (b) measured as a function of bias current for
sample \#5
at $T\approx 20$mK, and $f\approx 0$.  Note that at low bias
$V_x$ and $V_y$ are very similar}
\end{figure}

\subsection{Hall voltages}
\label{Hall voltage}
In an array which is strongly superconducting there is no Hall 
voltage\index{Hall voltage} since the Hall probes would be shorted. In an array with a 
strong Coulomb blockade the whole array is insulating and therefore 
the Hall probes are effectively disconnected, and no Hall voltage can be 
measured. Therefore it is not surprising that it is only the 
``intermediate'' arrays which show some Hall voltage.
In Fig.\,\ref{IV}a we see the $I_xV_x$ characteristics for sample \#5 which 
shows a sharp dip in the current before entering the flux-flow 
regime, characterized by $I$ being proportional to $V$. The $I_xV_y$ curve, shown 
in Fig.\,\ref{IV}b, displays a very similar feature, but the Hall 
voltage is about two orders of magnitude smaller.  As the current is 
reversed, the Hall voltage changes sign as expected.

In the flux-flow regime, both $V_x$ and $V_y$ increases linearly with 
applied current until the longitudinal voltage reaches the sum gap 
voltage of the entire array ($N\cdot 2\Delta_{0}\approx 63$\,mV).  At this 
point, the Hall voltage reaches a maximum of about 0.2\,mV and then 
starts to fluctuate, gradually decreasing almost to zero. At still 
higher currents, the Hall voltage displays rich structure, which can 
also be seen in the derivative of $V_x$ with respect to $I_x$.

At very high currents, when $V_x$ is on the normal resistance branch of 
the $I_x-V_x$ curve, the Hall voltage $V_y$ also increases linearly 
with the applied current, with slope $2.1\,\Omega$.  
Because this slope is a small fraction of $R_N$ (0.02\% for array \#5 
and 0.1\% for array \#7) we conclude that small non-uniformities 
exist in the array.  For array \#7, the Hall voltage as a function of 
bias current shows very similar behavior to that of array \#5.


\section{The insulating transition}
\label{Insulating}
\begin{figure}[b!]
\centerline{\epsfig{file=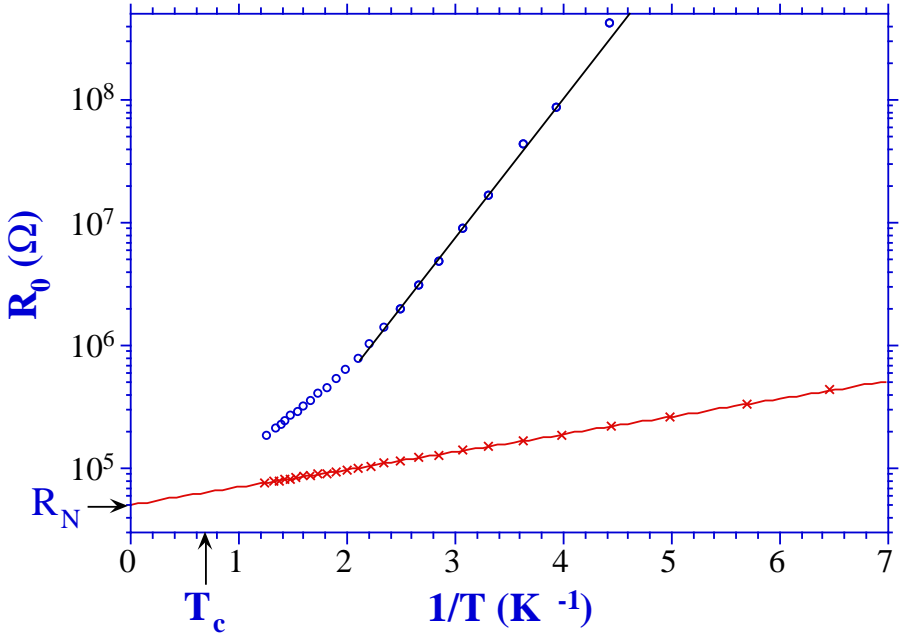,width=12cm}}
\vspace{10pt}
\caption{\label{Thermalactivation}The zero bias resistance $R_0$ for array 
\#10, vs. $1/T$ for $B=0$\,G, 
S-state ($\circ$), and for $B=1400$\,G, N-state ($\times$). Note the very good fit to the 
thermal activation Arrhenius dependence for the N-state. The resulting slope is very 
close to  $\frac{1}{4}E_C$ and the extrapolation to infinite temperature ends up right 
at $R_N$. At low temperature the data in the S-state can also be fitted to an Arrhenius 
dependence with a slope close to  $\frac{1}{4}E_C+\Delta_0$.}
\end{figure}

The zero bias resistance was measured as a function of both temperature 
and magnetic field. All arrays showed an "insulating" behavior in the N-state, 
meaning that $R_0$ increased as a function of decreasing temperature. Over a 
fairly wide range of temperature the $R_0(T)$ dependence was exponential as can 
be seen in Fig.\,\ref{Thermalactivation}, indicating
thermal activation\index{thermal activation} of charge solitons. 
However, $R_0$ saturates and is no longer temperature dependent at the lowest 
temperatures, (beyond the range displayed in Fig.\,\ref{Thermalactivation}). 
This saturation has been discussed in Ref.\cite{Delsing-PRB94} and 
here we will concentrate on the thermal activation.

In the S-state, $R_0$ increased even more rapidly as a function of decreasing 
temperature for the "insulating" arrays. The $R_0$ versus $T$ curves can be 
divided into three temperature regions (see Fig.\,\ref{Thermalactivation}). 
i) In a temperature range 
below 500\,mK, $R_0$ increases exponentially with decreasing temperature, and 
an activation energy can be defined. 
ii) At higher temperatures the 
dependence was not purely exponential due to the temperature dependence of 
the superconducting gap. 
iii) For the lowest temperatures $R_0$ became larger 
than 1\,G$\Omega$ (not shown in the figure), and could not be measured accurately 
with our experimental setup.

In a temperature interval roughly between 200\,mK and 500\,mK, $R_0$ for the 
"insulating" arrays could be fitted by a thermal activation dependence over the whole 
magnetic field range, such that

\begin{equation}
\label{Arrhenius}
R_0(B,T)= b\cdot \exp{\frac{E_a(B)}{k_BT}}
\end{equation}

where $E_a$ is the activation energy and $b$ is a constant.  It should be noted that 
there is a region of magnetic field slightly below $B_c$ where the superconducting 
gap goes to zero in the temperature interval where the fit is made to determine $E_a$. 
Therefore, the $ln(R_0)$ vs. $1/T$ plots are not perfectly linear at those magnetic 
fields. However the Arrhenius law (\ref{Arrhenius}) is still a fair approximation.

The fact that we do not observe a KTB charge unbinding transition in 
these arrays is not altogether surprising. It was shown by Zaikin and 
Panyukov \cite{Zaikin/Panyukov-LT21},
that the effect of offset charges\index{offset charges} will effectively cut of the 
logarithmic interaction between the charges. Also, the finite size of 
our samples is probably a limiting factor.

In the N-state, $E_a$ was found to be close to $\frac{1}{4}E_C$ (see 
Table \ref{Table1}), 
and b was very close to $R_N$, for all the "insulating" arrays. 
This value of $E_a$ agrees well with that of Tighe et al.\cite{Tighe-PRB93}. 
If the conduction is caused by thermal activation of 
single electron solitons we expect that $E_a=\frac{1}{4}E_C$ in the N-state. 
The energy required to create a soliton anti-soliton pair starting 
from an uncharged array, i.e. the tunneling of a single charge, 
is the so called core energy $E_{core}=e^2/4C=\frac{1}{2}E_C$ for single electrons, 
and four times as high $E_{core}=2E_C$ for 
Cooper-pairs. Similarly to thermal activation in other systems the activation energy 
becomes half of the core energy.

We next consider the S-state case, where we can imagine two alternative 
transport mechanisms. On one hand, we can calculate the energy needed to 
brake up a Cooper pair and to create a SES pair. We would 
expect $E_{core}=\frac{1}{2}E_C+2\Delta _0$ and therefore, 
$E_a=\frac{1}{4}E_C+\Delta _0$ if the charge transport was 
entirely due to SESs, created from broken Cooper pairs. If on the other hand, 
we assume that only CPSs are activated, we would expect the activation energy 
to be four times higher (because of the 2e charge) than for SESs in the N-state, 
so that $E_a=E_C$. This picture is of course a bit naive because the Josephson 
coupling energy is not taken into account. Nevertheless, if this simple picture 
holds, we would expect that for arrays with $\Delta _0/E_C<\frac{3}{4}$, 
thermal excitation of 
SESs would always be advantageous, and we would expect $E_a= E_C+\Delta (B)$. 
If on the other hand $\Delta _0/E_C>\frac{3}{4}$ , we would expect a crossover 
from activation of 
CPSs to activation of SESs as $\Delta $ is suppressed by the magnetic field. All the 
"insulating" arrays had $\Delta _0/E_C>\frac{3}{4}$, and we thus expect the crossover
behavior with magnetic field.

\begin{figure}[t!]
\centerline{\epsfig{file=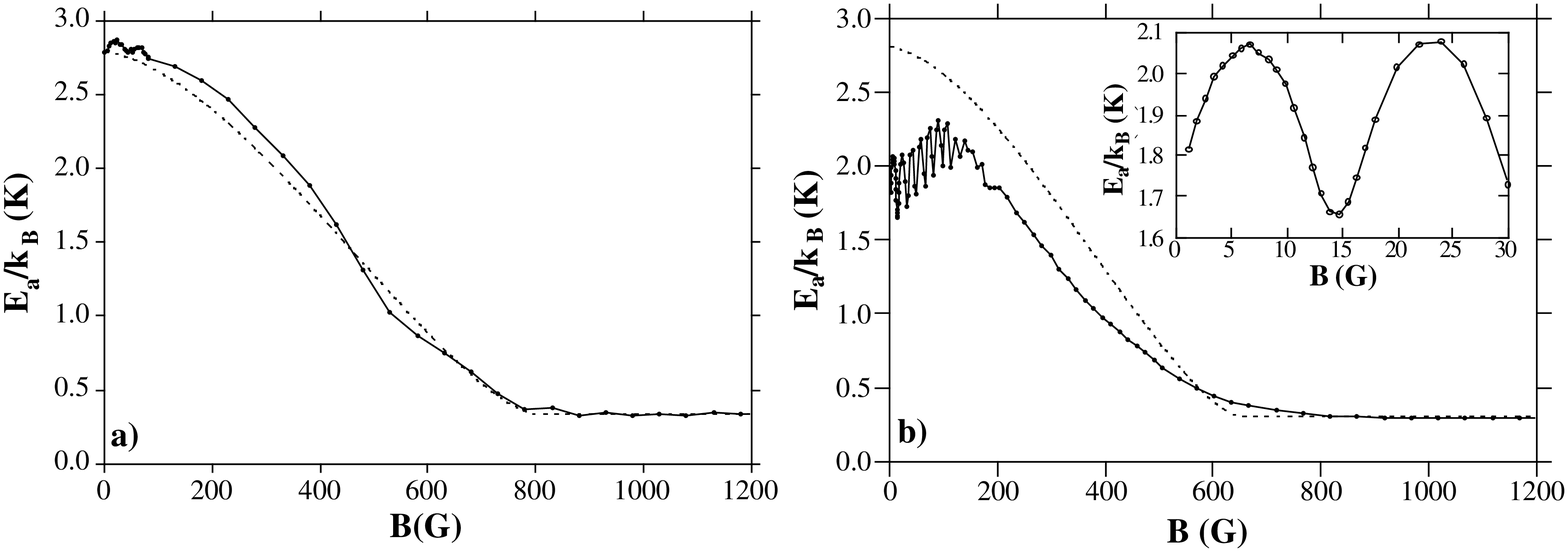,width=14cm}}
\vspace{10pt}
\caption{\label{Ea(B)}The activation energy $E_a$, as a function of magnetic
field $B$ for two of the arrays.
At high magnetic field (in the N-state), $E_a$ is close to $\frac{1}{4}E_C$. 
The dashed lines corresponds to $\frac{1}{4}E_C+\Delta (B)$. 
a) Array \#13. Note the good agreement between the experimental data 
and the dashed line. 
b) Array \#9. Note the oscillations, the period correspond 
to one flux quantum per unit cell, the amplitude is roughly equal to $E_J$.}
\end{figure}

Our observations suggest a somewhat more complicated picture. As we 
enter the S-state, by lowering the magnetic field below$B_c$, $E_a$ increases 
for all the "insulating" arrays. At zero magnetic field, we find that $E_a$ is equal to 
or smaller than  $\frac{1}{4}E_C+\Delta _0$ but larger than $E_C$, for all seven 
arrays. In Fig.\,\ref{Ea(B)} the magnetic field 
dependence of $E_a$ is shown for two of the arrays, and it is compared to
$\frac{1}{4}E_C+\Delta (B)$, which is represented by a dashed line. Here, $\Delta _0$ 
is measured for each array and the form of the function
$\Delta (B/B_c)/\Delta _0$ is also 
determined from the measurements \cite{Delsing-PRB94}.

From our simple picture outlined above we would expect the 
ratio $\Delta _0/E_C$ to be 
important. We find that for arrays with a ratio $\Delta _0/E_C$ less than 2, 
(arrays \#12,13, and 15), $E_a(B)$ has a behavior which is very close 
to $\frac{1}{4}E_C+\Delta (B)$, as can 
be seen for array \#13 in Fig.\,\ref{Ea(B)}a. Arrays \#12 and \#15 show a very similar 
behavior. Tighe et al.\cite{Tighe-PRB93} have also obtained the same result at $B=0$ 
for arrays where the ratio $\Delta _0/E_C$ was less than 2.
However,  for the other arrays \#9-11, and \#14, where $\Delta _0/E_C$ was larger 
than 2, $E_a(B)$ is lower than  $\frac{1}{4}E_C+\Delta (B)$. This can be seen 
in Fig.\,\ref{Ea(B)}b, for array \#9. It is obvious that the ratio $\Delta /E_C$ is 
important. We find that the critical value is about 2.

In several arrays, we observe oscillations in $E_a(B)$. The 
period of oscillation corresponds to one flux quantum per unit cell (see 
the inset of Fig.\,\ref{Ea(B)}b). The period agrees very well with the $B_0$ values, 
determined from geometry. The oscillations show 
that the Josephson coupling affects the activation energy. The oscillations are 
observed in all arrays where the measurements were taken with sufficiently 
small steps of the magnetic field. The amplitude of the oscillations is 
roughly $E_J$. Arrays with a large $E_J$ also showed an increase of 
(the average of) $E_a$ with increasing $B$ resulting in a peak at 
100 to 200\,G (Fig.\,\ref{Ea(B)}b).

These  effects can be understood since the creation of CPS/antiCPS pairs 
should be dependent on the Josephson coupling. For a weaker Josephson 
coupling, it should be harder to create CPS/anti-CPS pairs and $E_a$ should 
increase. The Josephson coupling $E_J\cos(\Phi/\Phi_0)$ is affected in two ways by the 
magnetic field. At low field the cosine part is affected, resulting in an 
oscillating $E_a$ with maxima where $(B=n+\frac {1}{2})B_0$, n being an integer. 
The oscillations of $E_a$ demonstrate clearly that at least part of the current at low 
bias is carried by Cooper-pair solitons. At higher magnetic field the 
increasing $E_a$ with increasing $B$ may be explained by a 
decreasing $\Delta$, and thereby also a decreasing $E_J$. At even higher fields 
$\Delta /E_C$ becomes smaller than 2 so that SES 
creation dominates, and thus $E_a$ decreases with increasing field 

In summary, our results for the N-state agree well with thermal activation 
behavior, and an activation energy of
$\frac{1}{4}E_C$ for single electron solitons can be 
extracted. For the S-state we find that as long as $\Delta /E_C<2$, pairs of SESs are 
created by breaking up Cooper pairs so that $E_a=\frac{1}{4}E_C+\Delta (B)$. For larger 
values, $\Delta /E_C>2$, pairs of CPSs are responsible for a substantial part of the 
charge transport and $E_{a}$ oscillates as a function of temperature.


\section{The magnetic field tuned superconductor insulator transition}
\label{SIT}

As mentioned previously arrays which are superconducting a low 
temperature can display a vortex-unbinding
KTB\cite{KT(B)} transition\index{KTB transition} 
to a resistive state above a certain critical temperature $T_{KTB}$.
In zero magnetic field, the theory for KTB vortex-unbinding gives a relation 
between the superconducting correlation length $\xi$ and the transition 
temperature $T_{KTB}$. The correlation length is determined by 
a control parameter which, for example, can be the disorder of 
the system. A Josephson junction array can be described by the 
classical 2D-XY model and can be associated to the KTB transition 
in continuous films \cite{LAT}. In the presence of disorder, the conductivity 
at low temperature is governed by variable-range hopping(VRH) 
\cite{Fischer-PRB89} and $\xi$ should scale with $T_{KTB}$. In a highly disordered 
film, the long-range vortex-pair order is destroyed, and $T_{KTB}$ 
is substantially suppressed compared with that of a disorder-free 
film.  Provided that the transition between insulator and superconductor 
is continuous, right at a critical disorder, $\xi$ should diverge 
and $T_{KTB}$ should vanish. If the disorder is smaller than but 
close to the critical disorder, Fisher \cite{Fischer-PRL90} predicted,
based on the analogy of VRH of vortices 
to VRH of electrons, that $\xi$ should diverge as ($B-B_{c})^{-\nu_B}$
with exponent $\nu_{B}\ge 2/d=1$, where d=2 is the dimension of 
the system. The scaling theory developed for the zero field 
case implies a power law dependence of $T_{KTB}$ on $\xi$, {\it i.e}.
$T_{KTB}\sim \xi^{-z_{B}}$, with an exponent $z_{B}$ of unity.
Furthermore, the dual transformation suggests 
a $T=0$, $B=B_{c}$ fixed point where the magnitude of the resistivity
(vector sum of the longitudinal and transverse components of the resistivity)
should also be universal and equal to $R_{Q}$.

\begin{figure}[t!]
\centerline{\epsfig{file=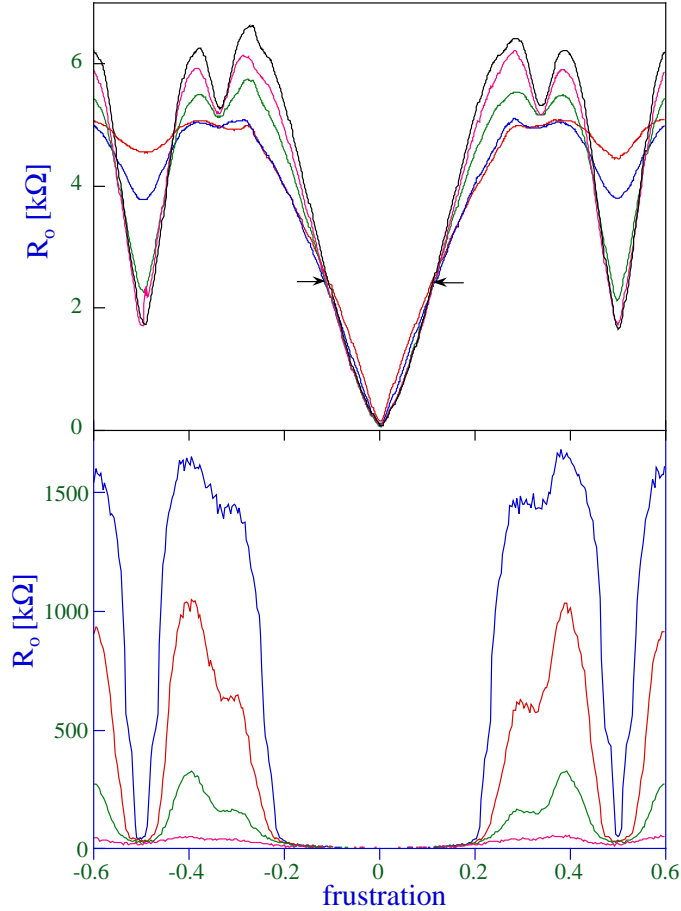,width=9cm}}
\vspace{10pt}
\caption{\label{Ro(f)}Linear resistance Ro at zero bias as a function of frustration for
a) array \#4 measured at several temperatures (from the top, 20, 200, 400, 600, 700 mK).
At $f=f_{c}\approx 0.12, dR_{0}/dT$ changes sign and, 
$R_{o}^{*} \approx 2.4\,$k$\Omega$.
b) array \#6 at low temperatures (from the top 50, 70, 90, 125 mK).}
\end{figure}

\begin{table}
\caption{\label{Table2}Parameters and scaling exponents for four of the arrays which showed the 
superconductor insulator transition. The variables are defined in the text.}
\vspace{0.3cm}
\begin{center}
\begin{tabular}{|c|c|c|c|c|c|c|}
\hline
\# & $E_{J}/E_{C}$ & $T_{KTB}$ & $f_{c}$ & {\it z}$_{B}\nu_{B}$ & $R_{0xx}^{*}$ & $R_{oxy}^{*}$ \\
 ~ &       ~       &  (K)   &   ~    &             ~          &  (k$\Omega$)  &   ($\Omega$)  \\
\hline
4 & 1.57 & 0.70$\pm$0.04 & 0.122$\pm$0.012 & 4.75$\pm$0.5 & 2.45$\pm$0.15 & - \\
5 & 1.24 & 0.46$\pm$0.04 & 0.039$\pm$0.007 & 8.20$\pm$1.0 & 1.23$\pm$0.10 & 28$\pm$5 \\
6 & 0.80 & 0.44$\pm$0.05 & 0.047$\pm$0.005 & 1.47$\pm$0.2 & 2.20$\pm$0.15 & - \\
7 & 0.90 & 0.35$\pm$0.04 & 0.034$\pm$0.010 & 4.45$\pm$0.5 & 1.61$\pm$0.15 & 34$\pm$5 \\
\hline
\end{tabular}
\end{center}
\end{table}

Experiments on homogeneous thin films in zero magnetic field 
demonstrated a SI transition\index{supreconductor insulator transition}
by changing the film thickness. 
The transition was found to occur when the film sheet resistance 
was close to a critical value of $R_{Q}$ \cite{Haviland-PRL89}.
A magnetic field tuned transition was found 
for films close to the critical resistance
\cite{Paalanen-PRL92,Hebard/Paalanen,Seidler,Tanda-PRL92,Wu/Adams} and agreement 
with the scaling theory \cite{Fischer-PRL90} was obtained.
In Josephson junction 
arrays, a SI transition can be achieved by changing the junction 
normal state resistance $R_{N}$ and the junction capacitance $C$
\cite{CCD-PhysicaScripta,Fischer-PRL90}. 
The transition is found to occur near a critical point at
$R_{N}\approx $R$_{Q}$ and $E_{J}/E_{C}\approx 1$.
We therefore expect a magnetic-field-tuned SI transition for 
arrays near the critical point.  However, there is a major difference 
between a uniform film and a regular array in the presence of 
an external magnetic field.  In the former case the vortices 
form an Abrikosov (triangular) lattice in the ground state \cite{Tinkham-book}
whereas in the latter case the vortices are pinned in a periodic 
potential, imposed by the array lattice \cite{Halsey}.  The array lattice 
results in a ground state energy which is an oscillatory function 
of the applied magnetic field with period $\Delta f$=1 \cite{Halsey}.
The ground state energy is not only periodic, 
but has minima at rational frustrations, {\it i.e.} $f$=1/2, 1/3, 2/3, etc.
From one point of view, frustration 
is simply proportional to magnetic field and should be related 
to $B$ in the scaling theory\cite{Fischer-PRL90}\index{scaling}.
From another point of view, 
frustration can be consider as introducing defects from the ordered 
lattice at rational $f$ values, in which case the KTB
transition under consideration 
is the melting of the vortex lattice.  We thus expect that the 
$f$ value can act as a control parameter for the SI transition, 
and the scaling theory can be applied, provided one accepts that 
the correlation length diverges at some critical value of
$f$ as $(f-f_{c})^{-\nu_{B}}$. Experimentally, we find that the SI transition 
occurs at $f$=n$\pm \delta$ and n+1/2$\pm \epsilon$ (where n is an integer and
$\delta , \epsilon <<1$), and that the scaling analysis is applicable 
to both cases.

The important parameters are listed in Table \ref{Table2}. The charging 
energies $E_{C}$ were judged from the offset voltage of the
$IV$-characteristics at large bias. $B_{o}$ 
was about 16 G for all the intermediate arrays.  For sample \#6 the superconducting 
mean-field-transition temperature, $T_{c}$=1.51\,K, was measured 
on an aluminum wire which was fabricated on the same chip as 
the array. The KTB transition temperature $T_{KTB}$ could be deduced 
from the onset of the linear dependence of $R_{o}(f)$ 
\cite{CCD-PRB96},
the values are listed in Table \ref{Table2}. This method 
of deducing $T_{KTB}$ has been confirmed both theoretically
\cite{KT(B),Halperin/Nelson} and experimentally
\cite{Hebard/Paalanen,Martin-PRL89}.

$R_{o}$($f$) for sample \#4 at $T<T_{KTB}$ is depicted in 
Fig.\,\ref{Ro(f)}. Below a critical frustration$f_{c }\approx$ 0.12, the resistance
is lower for lower temperature, indicating a superconducting transition. 
Above $f_{c}$, the resistance is higher for lower temperature, 
implying an insulating transition. The resistance at $f_{c}$, $R_{o}^{*}$,
can be identified from Fig.\,\ref{Ro(f)}a and is about 2.4 k$\Omega$.  

\begin{figure}[b!]
\centerline{\epsfig{file=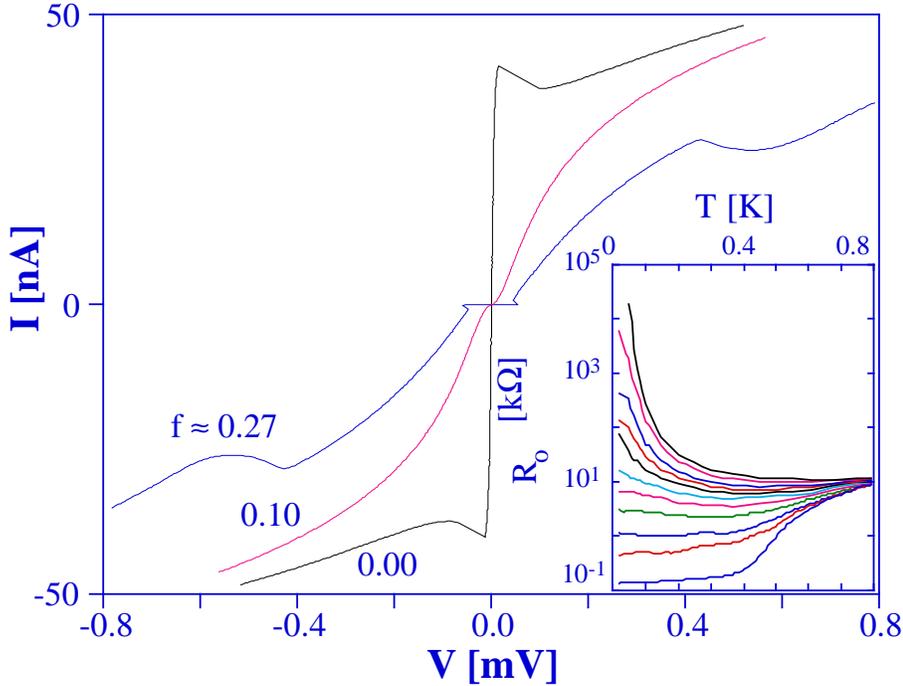,width=12cm}}
\vspace{10pt}
\caption{\label{SIT-IV}Magnetic field dependence of on the IV characteristics of array 
\#6 taken at 15\,mK.  In the superconducting state
($f\approx 0, R_{0}\approx 132\Omega$) it shows
a clear supercurrent whereas in the insulating state ($f\approx 0.27$, 
$R_{0}\approx $37.2\,M$\Omega$)
it exhibits a back-bending feature.  The zero bias resistances differ by
more than five orders of magnitude, while the change in magnetic field is
only 4.4\,G. The inset shows $R_{0}(T)$ for the same array for 0$<f<$0.27.}
\end{figure}

Fig.\,\ref{Ro(f)}b illustrates the $R_{o}(f)$ dependence for sample \#6 at
low temperatures.  While 
arrays \#4 and \#5 do not show a resistance higher than their normal 
resistances $R_{N}$ at any $f$ in the accessible temperature range,
the resistance 
of arrays \#6 and \#7 in the most insulating case
(at $f\approx 0.4$) are much greater than their $R_{N}$.
Remarkably, the resistance of array \#6 
changes by more than five orders of magnitude going from
$f\approx 0$ to $f\approx 0.27$ at 15\,mK.  The IV characteristics measured 
in the superconducting state and in the insulating state for 
array \#6 are displayed in Fig.\,\ref{SIT-IV}. In the superconducting state, 
the array exhibits clear Josephson-like current, with 
$R_{o}(f\approx 0)$
$\approx$ 130 $\Omega$, whereas in the insulating state, the 
array shows an insulating behavior with $R_{o}(f\approx $0.27) 
$>$37~M$\Omega$ and a back-bending feature in the IV characteristics 
similar to the behavior of higher resistance arrays
($R_{N}\approx$17 k$\Omega$) reported earlier 
\cite{GeerligsPhysicaB,CCD-PhysicaScripta}.
This feature is easily smeared by increasing temperature, 
resulting in a drastic decrease in the measured $R_{o}$ in the 
insulating phase as seen in Fig.\,\ref{Ro(f)}b. $R_{0}(T)$ curves for array 
\#6 in the range $0<f<0.27$ are shown as an inset of Fig.\,\ref{SIT-IV}.
The flattening-off 
of the resistance at non-zero frustrations at low temperatures 
is attributed to a finite size effect, explained within the context 
of the (vortex) VRH picture \cite{Fischer-PRL91}. In this picture, the hopping 
range increases at low temperature.  When the hopping length 
becomes larger than the sample size, a temperature-independent 
resistance is expected.  This flattening-off behavior was also 
reported in Refs.\cite{vdZant-PRL92,Delsing-PRB94} and can be shown to depend
on the sample size \cite{CCD-PRB96}.

\begin{figure}[b!]
\centerline{\epsfig{file=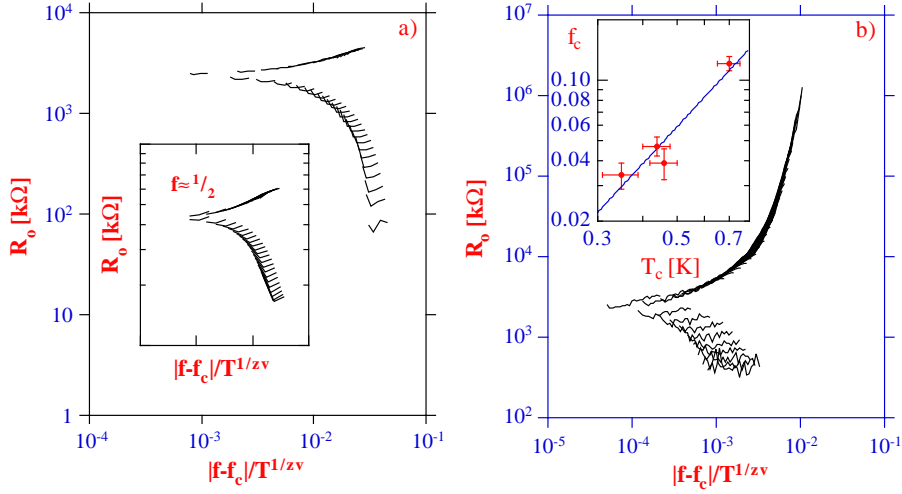,width=12cm}}
\vspace{10pt}
\caption{\label{Scaling}$R_{0}$ as a function of the scaling parameter
$|f-f_{c}|\cdot T^{-1/z_{B}\nu _{B}}$ for array \#4 and array \#6,
$0<f<0.2$. The data collapse onto one curve: the upper part for the insulating
transition and the lower part for the superconducting transition.
The low temperature points on the superconducting side deviate from the general
trend due to the finite array size.  a) Sample \#4.  The inset shows the scaling
for array \#4 close to full frustration, $0.5<f<0.6$.  b) Sample \#6.  The inset shows
a log-log plot of the critical frustrations $f_{c}$ as a function of the
Kosterlitz-Thouless transition temperature $T_{c}$ for the four measured samples.
The critical exponent $z_{B}\approx $1.05 can be obtained from the relation
$f_{c}\sim T_{KTB}^{2/z_{B}}$.}
\end{figure}

To appreciate the scaling\index{scaling} theory, all the data from both the 
superconducting and the insulating sides should collapse onto 
a single curve when plotting the resistances against the scaling 
variable $|f-f{c}|\cdot T^{-1/z_{B}\nu_{B}}$.
This is done using both $f_{c}$ and $z_{B}\nu_{B}$ as free parameters.
The curves were determined 
by minimization of the mean square deviation from an averaged 
curve on the insulating branch. The scaling was performed in 
the temperature range 50\,mK$<T<T_{KTB}$ and in the frustration 
range $0<f<0.2$. The scaling parameters as well as $R_{o}^{*}$
are listed in Table 1. Scaling curves for arrays \#4 and \#6 are 
shown in Fig.\,\ref{Scaling}. All data point in the insulating branch collapse 
onto the same trend.  For the superconductor transition the low 
temperature points deviate from the general trend due to the 
finite size effect.

\begin{figure}[t!]
\centerline{\epsfig{file=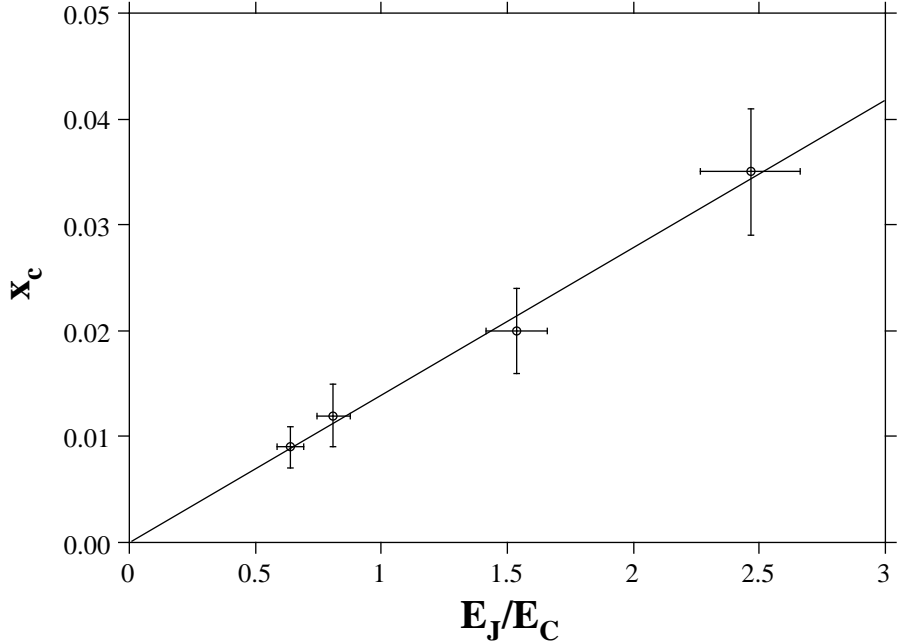,width=12cm}}
\vspace{10pt}
\caption{\label{XcvsEj/Ec}The critical $x$-value, $x_{c}$  
versus the $E_{J}/E_{C}$ value for the four arrays \#4-7. Our data shows 
a clear linear dependence which is not predicted by theory}
\end{figure}

The form of the scaling curves 
are very similar in the different cases, in fact it is possible 
to make the curves from all four arrays overlap by offsetting 
the graphs slightly in the $x$ and the $y$ direction. The different offsets in the $y$ 
direction demonstrates that the zero temperature fix point is different 
for the different arrays and that we get sample dependent values of 
$R_{0}^{*}$. The offset in the $x$-direction shows an interesting,
and unpredicted, linear 
dependence on the $E_{J}/E_{C}$ ratio. This is shown in 
Fig.\,\ref{XcvsEj/Ec} where $x_{c}$ is ploted versus the $E_{J}/E_{C}$ ratio.
$x_{c}$ is the $x$-value at which the 
superconducting transition extrapolates to zero resistance. 
Disregarding the low temperature 
points on the superconducting side, we find that we can describe all 
the data of the four arrays with a single formula 
(Eq.\,\ref{Scalingformula}), where the scaling 
parameter deduced from theory is normalized to the $E_{J}/E_{C}$ ratio.

\begin{equation}
\label{Scalingformula}
R_0(f,T)=R_{0}^{*}\cdot F( \frac{E_{C}}{E_{J}} \cdot \frac{|f-f_{c}|}{T^{1/z_{B}\nu_{B}}})
\end{equation}

Here F is a universal function describing both the insulating 
transition and the superconducting transition, and $R_{0}^{*}$ is a 
nonuniversal, sample dependent parameter.

We find, similar
to Hebard and Paalanen \cite{Hebard/Paalanen}, that the quality of the scaling 
is more sensitive to $f_{c}$ than to $z_{B}\nu_{B}$.
Using an improper $f_{c}$ value may cause large discrepancies at the left 
side of the plot and slightly shift $R_{o}^{*}$, whereas an 
improper $z_{B}\nu_{B}$ degrades the scaling but does not change 
$R_{o}$$^{*}$.  Note that the scaling analysis is limited to 
low $f$ values by the commensurability between the applied flux 
and the array lattice. The theory predicts that the transition 
temperature should scale with the correlation length as $T_{KTB}\sim \xi^{-z_{B}}$.
This, together with the fact that $\xi$ 
depends on the critical frustration as $f_{C}\sim \xi^{-2}$,
leads to the relation $f_{C}\sim T_{KTB}^{2/z_{B}}$.
Fitting $f_{c}$ and $T_{KTB}$ for the four samples as shown in the 
inset of Fig.\,\ref{Scaling}b, we can deduce $z_{B}=1.05$.
This can be compared to the theoretically predicted 
value of unity \cite{Fischer-PRL90}.

In the case of $f$=1/2 the region of sample parameters where the phase 
transition is observable shifts to lower $R_{N}$ and larger $E_{J}/E_{C}$.
For array \#4 we find $f_{c}=0.5\pm \epsilon$, with $\epsilon=0.055\pm0.001$ and $R_{o}^{*}$ 
is 4.33$\pm$0.15 k$\Omega$.  The scaling is shown as an inset 
in Fig.\,\ref{Scaling}a, the frustration range was $0.5<f<0.6$.
For array B, we find $\epsilon =0.030\pm 0.003$ and
$R_{o}^{*}\approx 5.40\pm 0.15\,$k$\Omega$.
Arrays \#6 and \#7 have a larger $R_{N}$ and a smaller
$E_{J}/E_{C}$ ratio, it is thus not surprising that the 
R$_{o}(f)$ curves at various temperatures do not cross near $f=1/2$.  

Returning to the S-I transition at $f$ close to zero, the scaling
curves for all four arrays 
exhibit similar bifurcation shape, but with a different $R_{o}^{*}$. 
As other measurements on 2D arrays \cite{vdZant-PRL92,vdZant-PRB95} and on superconducting 
films \cite{Seidler,Tanda-PRL92} also showed a different $R_{o}^{*}$, we conclude 
that $R_{o}^{*}$ is non-universal and sample dependent. 
According to theory \cite{Fischer-PRL90} the vector sum of the longitudinal resistance 
$R_{oxx}$ (previously $R_{o}$) and the Hall 
resistance $R_{oxy}$ should be universal and equal to R$_{Q}$ 
at $f=f_{c}$,  To check this prediction, we performed measurements 
of $R_{oxy}$ for arrays \#5 and \#7.  Four pairs of Hall 
voltage probes allow us to check the spread of the junction parameters 
in an array, which is found to be within 3\%.  $R_{oxy}(f)$ shows
a rich structure, the details will be published 
elsewhere.  For both arrays R{$_{oxy}$} is of the order of 30\,$\Omega$, 
see Table \ref{Table2}. 
The sum of $R_{oxx}^{*2}$ and $R_{oxy}^{*2}$ is thus smaller than 
$R_{Q}^{2}$for 
both arrays. It should be noted that the smallness of $R_{oxy}^{*}$
compared with $R_{oxx}^{*}$ agrees well with recent experiments on thin 
films \cite{Paalanen-PRL92}. The Hall angle at the transition, is about 
1.2$^o$ for both arrays.

In the thin film case \cite{Paalanen-PRL92} a critical field was found also 
for the $R_{oxy}$ data. In our case the linear region in 
the $I_{x}$ vs. $V_{y}$ characteristics is very small at finite 
$f$, which limits the excitation current and, consequently, 
the resolution in $R_{oxy}$. Therefore it is hard to determine the 
crossing point in the $R_{oxy}(f$) curves at $T<T_{c}$ for both array 
\#5 and \#7. Nevertheless, 
the frustration above which the $R_{oxy}(f)$ curves at various T start
to deviate from each other 
seems to be very close to $f_{c}$.  This is in contrast to the case
of disordered 
films \cite{Paalanen-PRL92}  where the ``critical field`` at which 
all $R_{oxy}(f)$
curves cross is higher than $B_{c}$ and is associated 
with the suppression $\Delta$. This is 
evidently not the case for 2D arrays, since the field needed 
for suppression of $\Delta$ of our arrays is about 800 G 
\cite{Delsing-PRB94}, which is much
greater than the critical field (=$f_{c}B_{0}$) of a few G.


\section{The Hall effect}
\label{Hall}
Superconducting films can, to some extent, be modeled as a 2D array 
of Josephson junctions, and understanding their transport behavior 
can be reduced to the problem of vortex dynamics.  Many interesting 
phenomena occurring in superconducting films can also be seen in 2D 
Josephson junction arrays.  In fact, the phenomena can be more easily 
modeled in the latter system because complications due to the (often 
unknown) microstructures do not exist, and phenomenological 
parameters such as the junction normal state resistance, $R_N,$ 
Josephson coupling energy $E_J$, and charging energy $E_C$, can be 
independently determined.  However, there is a major difference 
between a uniform film and a regular array in the presence of an 
external magnetic field.  In the former case the vortices form an 
Abrikosov (triangular) lattice in the ground state whereas in the 
latter case the vortices are pinned in the periodic potential imposed 
by the array lattice \cite{Halsey}. The array lattice of loops of area A, 
results in a ground state energy which is an oscillatory function of 
the frustration. The ground state energy is 
not only periodic, but has minima at rational frustrations, {\it i.e.} 
$f=1/2, 1/3, 2/3$, etc. \cite{Halsey}. In the vicinity of these rational 
frustrations, the dynamics is dominated by the motion of "defect" 
vortices \cite{Rzchowski-PRB94} and the vorticity of the majority defect is reversed 
upon passing through these rational frustrations.

\begin{figure}
\centerline{\epsfig{file=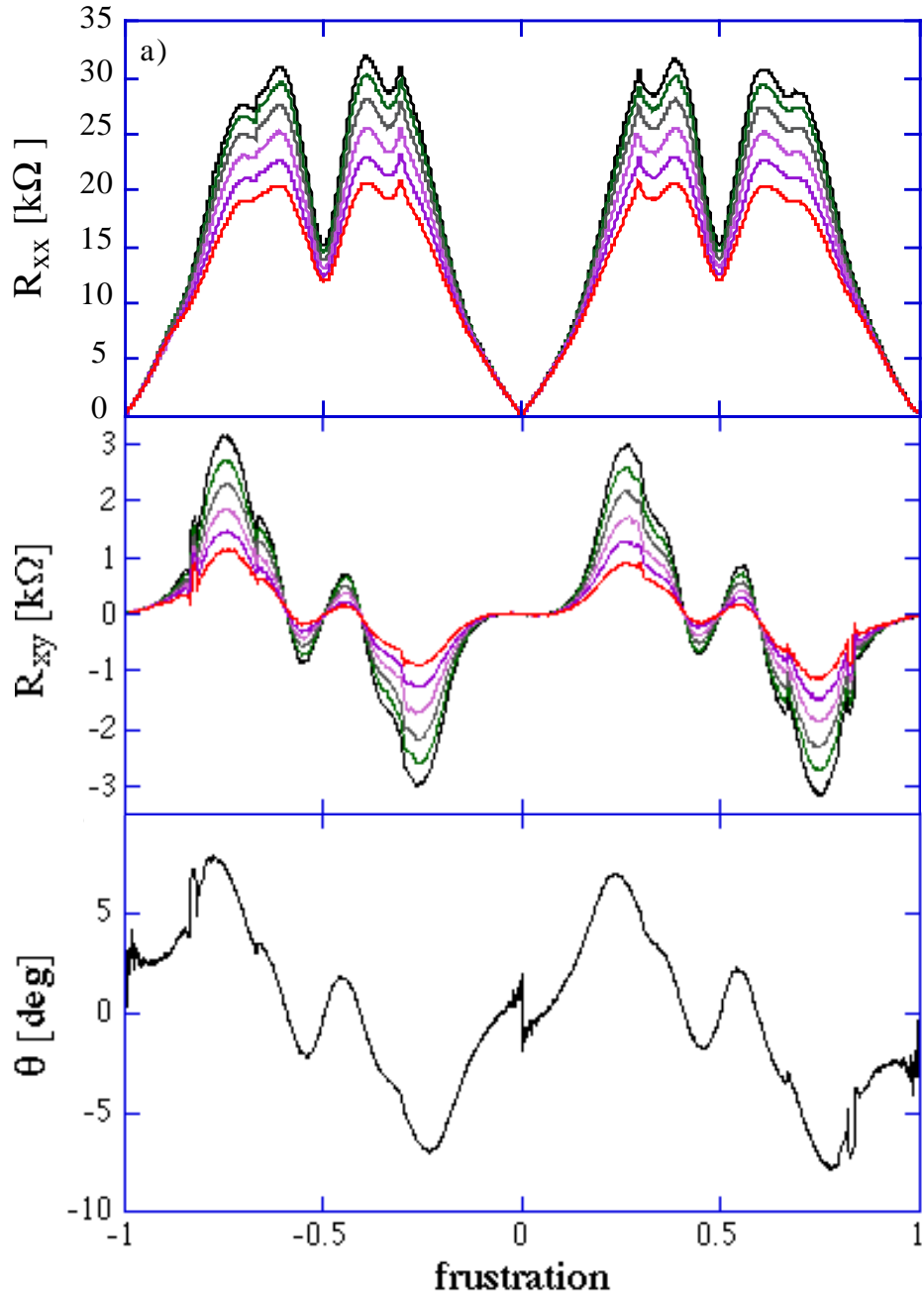,width=14cm}}
\vspace{10pt}
\caption{\label{Rvsf}(a) The longitudinal resistance $R_{0xx}$ and (b) the Hall 
resistance $R_{0xy}$, and (c) the Hall angle $\Theta$, as functions of 
frustration. $R_{0xx}$, and $R_{0xy}$ are shown for various 
temperatures (from the top at $f\approx$0.25, $T=$ 20, 75, 100, 125, 
150, 175\,mK). $R_{0xx}$ is symmetric at $f=0$ and $f=\pm 1/2$ 
whereas $R_{0xy}$ changes sign upon passing through these 
frustrations. The data are for sample \#7}
\end{figure}

The frustration dependence of $R_{0xx}$ and $R_{xy}$ for sample \#7 at 
various temperatures is shown in Fig.\,\ref{Rvsf}.  $R_{xx}$ is an 
oscillatory function of the applied magnetic field with period 
$\Delta f=1$ and minima 
at rational frustrations,  $f=1/2, 1/3$ and 2/3 as can be seen in 
Fig.\,\ref{Rvsf}.  There is a critical frustration, $f_c\approx 
0.034$,  below which the resistance decreases as the temperature is 
lowered, and above which the resistance 
increases as the temperature is lowered. The 
longitudinal resistance at $f_c$, $R_{0xx}^{*}=1.61\,$k$\Omega$ , can 
be identified from an expanded view of Fig.\,\ref{Rvsf}a around $f=f_c$, as 
well from a scaling analysis on these curves which we have 
discussed in detail in Section \ref{SIT}, see also 
Ref.\cite{CCD-PRB95}.

The Hall resistance\index{Hall resistance}
$R_{0xy}$ as a function of frustration is shown in 
Fig.\,\ref{Rvsf} where we see that it also oscillates with the 
applied field having the same period $\Delta f=1$.  At $f=0$, 
$R_{0xy}$ is zero, and $R_{0xy}(f)$ has a very small negative slope.  
As the frustration is increased, $R_{0xy}$ goes through a minimum, 
increasing to $R_{0xy}^{*}= 34\,\Omega$ at $f=f_c$. Thereafter it 
rapidly increases, reaching a  maximum value at $f\approx 0.23$.  As 
the frustration is increased further, $R_{0xy}$ starts to decrease and 
at $f>2/5$, it becomes negative. At $f\approx 0.45$, $R_{0xy}$ 
reaches a local minimum and starts to increase to zero at $f=1/2$.  
$R_{0xy}(f)$ is anti-symmetric about $f=0$ and locally anti-symmetric 
about$f=1/2$. In the raw data a small symmetric part arises due to 
sample non-uniformities. This symmetric part can be removed by 
taking $(R_{0xy}(f)-R_{0xy}(-f))/2$.  The removed symmetric part looks 
identical to, but is only 3\% of, $R_{0xx}$.  The data shown in 
Fig.\,\ref{Rvsf} has the symmetric part removed.  The difference in 
shape at $f=0$ and $f=\pm1$ is not understood although it is 
reproducible for $-1.2<f<+1.2$.

The $R_{0xx}$ and $R_{0xy}$ data can be combined to generate a third 
plot of the Hall angle $\Theta \equiv arctan(R_{0xy}/R_{0xx})$ which is shown 
in Fig.\,\ref{Rvsf} at $T=20$\,mK. Comparing our results to those of 
van Wees et al. \cite{Wees-PRB87} we find a much larger
Hall angle\index{Hall angle}.  Furthermore 
they did not observe an anti-symmetric $R_{0xy}$ versus $f$ curve.  
This is probably because their array was in the classical limit 
$E_J\gg E_C$, whereas ours were in the quantum limit $E_J\approx 
E_C$ where a finite $R_{0xy}$ has been predicted \cite{Fischer-PRL90,Otterlo-PRB93}

\begin{figure}[b!]
\centerline{\epsfig{file=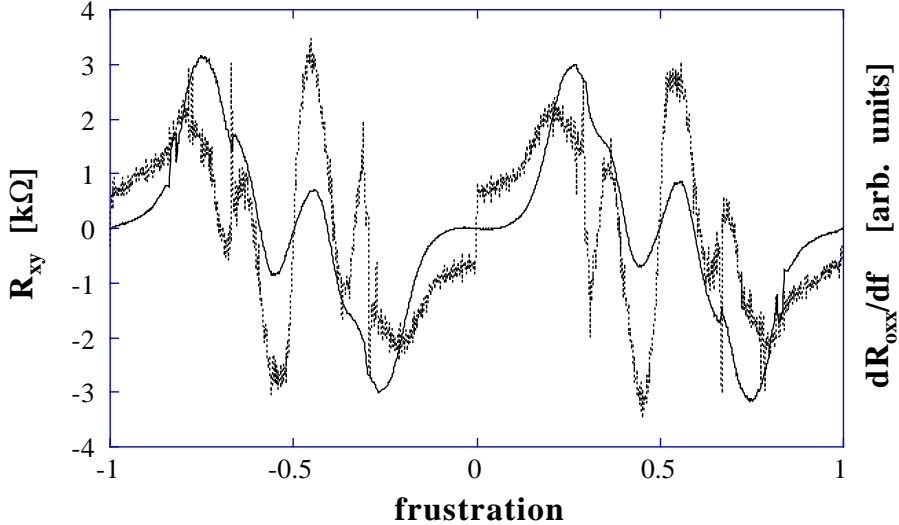,width=12cm}}
\vspace{10pt}
\caption{\label{Deriv}A comparison of $R_{0xy}$ (solid curve) and $dR_{0xx}/df$ 
(dotted curve) for sample \#7 at $T=20$mK, plotted with arbitrary 
scale.  Both curves show a similar behavior, minima and maxima occur 
at the same values of frustration}
\end{figure}

The Hall angle can be interpreted as the angle between the vortex 
velocity vector $\vec V_v$. and a unit vector perpendicular to the 
direction of current flow $\vec J$.  The moving vortices generates an 
electric field given by

\begin{equation}
\label{E-field}
\vec E=-q_vn_v\phi_0\vec V_v\times \hat z
\end{equation}

with $q_v=\pm1$ describing the vorticity, $n_v$ the area density of 
vortices, and $\hat z$ the unit vector perpendicular to the plane of 
the array.  Vortex motion in the direction of $J\times z$ creates a 
field $E$ parallel to the transport current $J$.  The force on a 
moving vortex is the Magnus force \cite{Ao/Thouless-PRL93}

\begin{equation}
\label{Magnus}
\vec F=n_se\phi_0\left(\vec V_s-\vec V_v\right)\times \hat z
\end{equation}

where $n_s$ is the superfluid electron density and
$\vec V_s = \vec J/n_se$ is the superfluid velocity. Due to this force, there is a 
component of vortex motion parallel to the applied current which 
produces a transverse field in the $-\vec J\times \hat z$ direction.

In two dimensional regular arrays, the field $\vec E$ at irrational 
frustrations is generated by the motion of "defect" or vacancy 
vortices.  From Eq.\,\ref{E-field}, it is clear that the sign change in the 
vorticity is responsible for the sign reversal of $R_{0xy}$ in the 
vicinity of $f=1/2$, because the defects have opposite vorticity to 
the field induced vortices.

We find that the structures in  $R_{0xy}(f)$ can be correlated to 
structure $dR_{0xx}/df$ as shown in Fig.\,\ref{Deriv}. The
correlation is most clear in the vicinity of $f=1/2$. This is similar 
to what is found in the Quantum Hall effect in two dimensional 
electron gases (2DEGs), where $R_{0xx}(B)\propto dR_{0xy}/dB$.
That we find traces of the opposite derivative law is probably related 
to the fact that our system is described in termes of vortices rather 
than in terms of electrons as the QHE in 2DEGs.


\section{Conclusions}
\label{Conclusions}

Based on the parameters of the tunnel junction we find that we can 
divide the properties of the 2D arrays into three different 
categories: i) Those which are dominated by the Coulomb blockade and 
go insulating at low temperature, ii) those which are dominated by the 
Josephson effect and go superconducting at low temperature, and iii) 
the intermediate arrays which just barely become superconducting at 
low temperature, but can be made insulating by applying a magnetic 
field.

We find that the transition for the insulating arrays can be well described by 
thermal activation  of charge solitons both when the electrodes are normal and 
when they are superconducting.
When the electrodes are in the normal state we find thermal activation of 
single electron solitons, with an activation energy $E_a\approx \frac{1}{4}E_C$. 
When the electrodes are in the superconducting state we find a much 
larger $E_a$. For arrays with $\Delta _0/E_C<2$, the activation energy is simply 
$\frac{1}{4}E_C+\Delta (B)$ indicating that Cooper pairs are broken up and that pairs of single 
electron solitons are created. When $\Delta _0/E_C>2$, the activation energy oscillates 
with $B$ at low magnetic field, demonstrating that Cooper pair solitons are 
created. The amplitude of these oscillations is roughly equal to $E_J$ and the 
period corresponds to one flux quantum (h/2e) per unit cell.

The threshold voltage for the insulating arrays also oscillates at low magnetic field demonstrating 
that Cooper pair solitons are injected at low field. For increasing magnetic 
field the average threshold voltage increases, due to the decreasing $E_J$. In the 
region 250 to 450\,Gauss we observe a peak in the threshold voltage which is 
interpreted as a crossover from Cooper pair soliton injection to single electron 
soliton injection.

We have observed a frustration-tuned superconductor-insulator 
phase transition in several of the ``intermediate'' arrays.  
A small applied magnetic field of 4.4\,G can change the zero-bias 
resistance of an array by more than 5 orders of magnitude.  We 
show scaling curves for both $f$=0 and $f$=1/2.  The results for our
four samples show a dynamic 
critical exponent of 1.05, in good agreement with the theory 
of the field-tuned S-I transition.
Our data indicate a sample dependent $R_{0}^{*}$.
Moreover, we have measured 
the Hall resistance at $f_{c}$, which is much smaller than $R_{Q}$.

For frustration values larger than the critical value the Hall resistance
is substantially larger and has a rich structure as a function of applied magnetic 
field. Reversal of the sign of the Hall 
resistance appears at several frustrations, which can be attributed 
to the change of sign of the "defect" vortices.  We find that the 
structure in $R_{0xy}(f)$ is similar to the derivative 
$dR_{0xx}(f)/df$.

\section*{Acknowledgements}
\label{Ackn}
We gratefully acknowledge fruitful discussions with S.M. Girvin, M. Jonson, 
J. E. Mooij, A. A. Odintsov, R. Shekhter, G. Sch\"on, A. Stern, M. Tinkham and
H. van der Zant.
Our samples were made in the Swedish Nanometer Laboratory. We would also like to 
acknowledge the financial support from the Swedish SSF, TFR and NFR, 
and the Wallenberg Foundation.


\begin{references}
\bibitem{quasiSchmid}
Strictly speaking this is not a Schmid Diagram since there one plots 
the damping term, which is determined by the tunneling resistance. 
For energies below the gap energy this is different from $R_{N}$.
\end{references}
\end{document}